# Estimation Of The Mass And The Spin Of The Neutron Star 4U 1728-34 Using Triples Of X-Ray Quasi-Periodic Oscillations

Radostina P. Tasheva[a)] and Ivan Zh. Stefanov[b)]

*Technical University of Sofia, Faculty of Applied Mathematics and Informatics, Department of Applied Physics, 8, Kliment Ohridski Blvd., 1000 Sofia, Bulgaria*

[a)] rpt@tu-sofia.bg
[b)]Corresponding author: izhivkov@tu-sofia.bg

**Abstract.** X-ray quasi-periodic oscillations (QPOs) were reported for 19 different observations of the atoll source 4U 1728-34. Three of them reveal triplets of QPOs: a lower and a upper kHz QPOs and a low-frequency QPO. Even though they are discernible only after averaging we assume them to be simultaneous. We apply a separate model (formula) for each of the three frequencies of the observed QPOs and compose a system which contains three algebraic equations for the three unknown parameters - the mass and the specific angular momentum of the central neutron star, and the radius of the orbit of orgin of the given triplet of QPOs, which allows us to find the numerical values of the later. Error propagation formulas are applied for the obtaining of interval estimates of the parameters of the star. This is done through implicit function differentiation. The calculations are repeated with several different models for the kHz QPOs.

## INTRODUCTION

Neutron stars are remnants of relatively massive stars (approximately 20 to 30 solar masses) that have suffered a severe supernova explosion. After the explosion thermonuclear reactions in the stars core stop and the star begins to collapse under the force of gravity. After the pressure of the neutron gas obtained when most of the protons and electrons of the star content coalesce is balanced by pressure due to gravity the object reaches its stable thermal equilibrium. It has lost a great amount of its mass during the explosion and is now approximately between one and two solar masses.With accordance of the law of conservarion of angular momentum the drastic decrease in the mass and respectively the moment of inertia leads to a great increase of its angular velocity corresponding to hundred revolutions per second. Such objects ususaly emit massive amount of energy in the X-ray part of the spectrum.This “primary” neutron star usually dies as a result of its period lengthening.

Neutron stars in binary systems can be revived by accreting matter and angular momentum from its companion. This type of objects are called recycled pulsars [1]. Eventually, after a time interval of speeding up the star reaches an equilibtium period of rotation which parameters are closely related to magnetic field strength and the rate of accretion.To this relatively stable period correspond quasiperiodic oscilations of its X-ray emission (QPO) [2]. They are expessed as intensity peaks superimposed on the continuous emission spectrum. The orirgin of the peaks can be traced to the movement of accretted hot matter as well as to the interaction between the angular moment of the star and the accretion disc that surrounds it. The models related to them will be discussed in the text.

The frequencies of the QPOs have a direct connection with the mass *M* and specific angular momentum *a* of the neutron star [3] .They are classified as low (LF) and high (HF). The LF ocssilation are variable and have frequencies up to 100 Hz. High frequent QPOs very often appear in pairs – lower - $\nu_L$ and upper $\nu_U$ which come usually related as $\nu_U : \nu_L$ =3:2.

The aim of our investigation is using data from the X-ray power-density spectrum of the atoll source 4U 1728-34 to assess the mass and the specific angular momentum of the central neutron star. As to do this we use several triplets of simultaneously observed quasi-periodic oscillations. The estimates for the mass and the specific angular momentum are given by the solution of a system of nonlinear algebraic equations. To evaluate the errors of the obtained estimates we use error propagation formulas.

## THE ATOLL SOURCE 4U 1728 -34

Zhang et al. [4] published for the fist time rough data from the space laboratory Rossy X-ray Timing Explorer concerning the atoll source 4U 1728-34. These data are consequently processesed by van Straaten et al [5] and published during 2002. For the the description of the power density spectra in the range between 0,01 and 1000Hz are used form 1 to 6 lorenzians. The results in the paper are presented for 19 time intervals. Only for 3 of them – time intervals 10,11 and 12 all three frequencies – the low $\nu_{LF_0}$ and the lower $\nu_{L_0}$ and upper $\nu_{U_0}$ high frequencies exist. We will assume that they appear simultaneously.The data for these frequiencies are given in the Table 1

**TABLE 1.** observational frequencies for the atoll source 4U 1728-34. Only data for the time intervals 10, 11 and 12 are given where all three frequencies are present.

| interval | $\nu_{LF_0}$, (Hz) | $\nu_{L_0}$, (Hz) | $\nu_{U_0}$, (Hz) |
|---|---|---|---|
| 10 | $42.14 \pm 0.77$ | $513 \pm 18$ | $849.5 \pm 2.0$ |
| 11 | $45.52 \pm 0.66$ | $561 \pm 11$ | $875.7 \pm 1.6$ |
| 12 | $46.70 \pm 0.91$ | $604 \pm 14$ | $907.6 \pm 2.5$ |

## FREQUENCY MODELING

The decription of the behavior of a test particle representing a lump of hot matter that originates from the companion star and is probably responsible for the creation of the peaks in the X-ray spectrum is convenient to be done in a polar coordinate system.The three geodesic frequencies- orbital $\nu_\Phi$, radial $\nu_r$ and vertical $\nu_\theta$ as well as their simple combinations can be used to to calculate the observational once. The equations expressing geodesical frequencies are taken from the paper of Aliev et al[3]. The low frequency has a value that links it to the Lense-Thirring precession: $\nu_{LF} = \nu_{nod} = \nu_\Phi - \nu_\Theta$. It is assumed to be a result of interaction between the orbits of the netron star itself and the accretion disk that surrounds it. The lower $\nu_L$ and the upper $\nu_U$ frequencies vary with correspondance of the chosen model.

The total precession model TP and TP1 are firstly proposed by Stuchlık et.al. [6] and introduce the term total precession period. This is the time for which a test particle moving along a quasieliptic orbit with a certain tilt and situated as close to star as possible will make one revolution in a such manner that its tilt and position with respect to the star will be the same. For these models the lower frequency will be $\nu_L = \nu_\theta - \nu_r$, and the upper frequency is $\nu_\theta$ for the TP model and $\nu_\Phi$ for the TP1 model.

The warped disc model WD[7] considers accretion dics warping due to the tilt of the accretion dics axis with respect to the star axis and the consecutive frequency superposition. The lower frequency will be $\nu_L = 2(\nu_\Phi - \nu_r)$, and the upper $\nu_U = 2\nu_\Phi - \nu_r$.

The tidal disruption model TD [8] assume that the lower and the uppere frequencies are $\nu_L = \nu_\Phi$ and $\nu_U = \nu_\Phi + \nu_r$ and their formation are a result of tidal disruption of a hot "blob" in the extremely strong gravitational field of the neutron star.

## FREQUENCY EQUATIONS AND THE ERROR PROPAGATION METHOD

We designate the observational values of the low frequency and the lower and the upper high frequencies as $\nu_{LF_0}$ $\nu_{L_0}$ and $\nu_{U_0}$, respectively, while $\nu_{LF}, \nu_L$ and $\nu_U$ are the theoreticaly derived ones. The non-linear algebraic equation that connect them depend on three independent parameters – the specific angular momentum *a*, the mass of the cenrtal object *M* and the specific radius *r* where corresponding frequencies originate:

$$f_1 \equiv \nu_{LF}(M,a,r) - \nu_{LF_0} = 0 \tag{1}$$
$$f_2 \equiv \nu_L(M,a,r) - \nu_{L_0} = 0$$
$$f_3 \equiv \nu_U(M,a,r) - \nu_{U_0} = 0$$

Therefore the free parameters are implicit function of the observed frequencies:

$$a = a(\nu_{LF_0}, \nu_{L_0}, \nu_{U_0}) \tag{2}$$
$$M = M(\nu_{LF_0}, \nu_{L_0}, \nu_{U_0})$$
$$r = r(\nu_{LF_0}, \nu_{L_0}, \nu_{U_0})$$

In order to assess the errors of the parameters we apply the following error propagation formulas

$$\sigma_a^2 = \left(\frac{\partial a}{\partial \nu_{U_0}}\right)^2 \sigma_{\nu_{U_0}}^2 + \left(\frac{\partial a}{\partial \nu_{L_0}}\right)^2 \sigma_{\nu_{L_0}}^2 + \left(\frac{\partial a}{\partial \nu_{LF_0}}\right)^2 \sigma_{\nu_{LF_0}}^2 \tag{3}$$
$$\sigma_M^2 = \left(\frac{\partial M}{\partial \nu_{U_0}}\right)^2 \sigma_{\nu_{U_0}}^2 + \left(\frac{\partial M}{\partial \nu_{L_0}}\right)^2 \sigma_{\nu_{L_0}}^2 + \left(\frac{\partial M}{\partial \nu_{LF_0}}\right)^2 \sigma_{\nu_{LF_0}}^2$$

Here the rules for implicit function differentiation are used for the calculation of the partial derivatives. In explicit form, one of these derivative, for example, is

$$\frac{\partial a}{\partial \nu_{U_0}} = -\frac{D(f_1, f_2, f_3)}{D(r, M, \nu_{U_0})} \cdot \left[\frac{D(f_1, f_2, f_3)}{D(r, M, a)}\right]^{-1} \tag{4}$$

where the corresponding Jacobi determinants

$$\frac{D(f_1, f_2, f_3)}{D(r, M, a)} = \begin{vmatrix} \frac{\partial f_1}{\partial r} & \frac{\partial f_1}{\partial M} & \frac{\partial f_1}{\partial a} \\ \frac{\partial f_2}{\partial r} & \frac{\partial f_2}{\partial M} & \frac{\partial f_2}{\partial a} \\ \frac{\partial f_3}{\partial r} & \frac{\partial f_3}{\partial M} & \frac{\partial f_3}{\partial a} \end{vmatrix} \quad \text{and} \quad \frac{D(f_1, f_2, f_3)}{D(r, M, \nu_{U_0})} = \begin{vmatrix} \frac{\partial f_1}{\partial r} & \frac{\partial f_1}{\partial M} & \frac{\partial f_1}{\partial \nu_{U_0}} \\ \frac{\partial f_2}{\partial r} & \frac{\partial f_2}{\partial M} & \frac{\partial f_2}{\partial \nu_{U_0}} \\ \frac{\partial f_3}{\partial r} & \frac{\partial f_3}{\partial M} & \frac{\partial f_3}{\partial \nu_{U_0}} \end{vmatrix} \tag{5}$$

occur. The rest of the partial derivatives in (3) are calculatetd in a similar manner. The standard errors of the observed frequencies, namely $\sigma_{\nu_{U_0}}, \sigma_{\nu_{L_0}}, \sigma_{\nu_{LF_0}}$ are given in Table 1.

The best estimate for the mass as well as its stantandart errors is given by its weighed average over the three independent measurements:

$$M_{\mathrm{av}} = \frac{\sum M_i / \sigma_i^2}{\sum 1/\sigma_i^2} \qquad \sigma^2_{\mathrm{av}} = \frac{1}{\sum 1/\sigma_i^2} \qquad (6)$$

Here $M_{\mathrm{av}}$ is the weighed mass and $\sigma^2_{\mathrm{av}}$ is the standard deviation. For the specific angular momentum we use similar formulas.

## MASS AND SPECIFIC ANGULAR MOMENTUM OF THE ATOLL SOURCE 4U 1728 -34

The mass and the angular momentum are calculated for each of the time intervals 10,11 and 12. The results for the TP,TP1 TD and WD model are presented in Tables from 3 to 6 respectively. The mass is calculated in solar mases and the angular momentum has no dimension.

**TABLE 3.** Mass and angular momentum of 4U 1728-34 according to the TP model

| interval | *M*, solar masses | *a*, dimensionless |
|---|---|---|
| 10 | $2.60 \pm 0.06$ | $0.35 \pm 0.01$ |
| 11 | $2.64 \pm 0.03$ | $0.35 \pm 0.01$ |
| 12 | $2.58 \pm 0.04$ | $0.34 \pm 0.01$ |

**TABLE 4.** Mass and angular momentum of 4U 1728-34 according to the TP1 model

| interval | *M*, solar masses | *a*, dimensionless |
|---|---|---|
| 10 | $2.85 \pm 0.06$ | $0.35 \pm 0.01$ |
| 11 | $2.90 \pm 0.03$ | $0.35 \pm 0.01$ |
| 12 | $2.88 \pm 0.04$ | $0.33 \pm 0.01$ |

**TABLE 5.** Mass and angular momentum of 4U 1728-34 according to the TD model

| interval | *M*, solar masses | *a*, dimensionless |
|---|---|---|
| 10 | $3.7 \pm 0.2$ | $0.85 \pm 0.14$ |
| 11 | $3.96 \pm 0.12$ | $0.68 \pm 0.08$ |
| 12 | $3.89 \pm 0.07$ | $0.61 \pm 0.05$ |

**TABLE 6.** Mass and angular momentum of 4U 1728-34 according to the WD model

| interval | $M$, solar masses | $a$, dimensionless |
|---|---|---|
| 10 | $3.5 \pm 0.2$ | $0.64 \pm 0.02$ |
| 11 | $3.80 \pm 0.13$ | $0.62 \pm 0.01$ |
| 12 | $3.86 \pm 0.16$ | $0.60 \pm 0.01$ |

The weighed average mass and angular momentum for all models are given in the Table 7.

**TABLE 7.** Weighed mass and angular momentum of 4U 1728-34 for all models

| model | $M_{av}$ , solar masses | $a_{av}$, dimensionless |
|---|---|---|
| TP | $2.62 \pm 0.02$ | $0.35 \pm 0.01$ |
| TP1 | $2.89 \pm 0.02$ | $0.35 \pm 0.01$ |
| TD | $3.89 \pm 0.06$ | $0.61 \pm 0.04$ |
| WD | $3.76 \pm 0.09$ | $0.61 \pm 0.01$ |

Using the method of Bambi [9] for calculation of the standard error Tasheva and Stefanov [10] reach similar results in the cases of relativistic pressesion models RP, RP1 and RP2.

As it is seen in the Tables from 3 to 6, estimates for the mass vary significantly . TP and TP1 models show mass consisitent with the acknowledged boundary for a neutron star – 2.62 and 2.89 solar masses respectively and is near this boundary – around 3 solar masses. The results are lower for the TP model.The TD and WD model shows results exceeding this limit – between 3.90 and 3.76 solar masses respectively.

Estimates for the specific angular momentum also can be divided into two groups. TP and TP1 models show specific angular momenta consistent with the Kerr metrics a≤0.4 [11]. The warped disk model WD and The TD model exhibit $a = 0.61$.

According to Table 7 TP and TP1 models predict appropriate mass lower than 3 solar masses and same average specific momentum values. The TD and WD models exibit average mass high for a neutron star and same specific momentum values.

## CONCLUSION

The processed by van Straaten et al [5] data concerning low (LF) and the two high frequencies - lower (L) and higher (H) of the peak values found in the power density X-ray spectra of the atoll source 4U 1728-34 are a good foundation for an assessment of their essential characteristics – the mass of the neutron star $M$ and its specific angular momentum $a$.The procedure of their extraction is based on the solving of a system of three non-linear algebraic equation where they are free parameters. The fundamental geodesic frequencies describing theoretical frequencies obey equations (3) and are combined in four models, namely TP,TP1,TD and WD.

The results can be divided into two groups. In the first group are the total precession models TP and TP1. They show specific angular momentum values consistent with Kerr metrics (a≤0.4) and masses near the upper boundary for a neutron star but they do not exceed it. In the second group are the tidal disruption model TD and warped disc model WD. They both show values for the specific angular momentum a≥0.4 and exibit average mass high for a neutron star.

## ACKNOWLEDGMENT

The authors would like to thank the Research and Development Sector at the Technical University of Sofia for the financial support.

## REFERENCES

1. Srinivasan, G. Recycled pulsars. New astronomy Reviews, Volume 54, Isssue 3-6, pp.93-100, 2010
2. Ronald A. Remillard R.A., Edward H. Morgan E.H., McClintock J.E., Bailyn C.D., Jerome A. Orosz J.A. & Greiner J., in: arXiv astro-ph 9705064Yan Cheung. CrossBar. Department of Computer Science, Emory University, 1997.
3. Aliev A. N., Esmer G. D., Talazan P., 2013, Class. Quantum Grav. 30, 2013
4. Zhang, W., Giles, A. B., Jagoda, K., Soong, J., Swank, J.H. and Morgan H.E. 1993, : Proc SPIE, 324, 2006
5. van Straaten S., van der Klis M, di Salvo T., Belloni T., Ap J, 568, 912 , 2002
6. Stuchlık, Z. , Torok G., Bakala P., A&A, in : arXiv:0704.2318, 2007
7. Kato S., 2005, PASJ,57,679
8. Cadez A., Galvani H., Kostic U., 2008, A&A, 487, 527
9. Bambi C., 2015, Eur. Phys. J. C 75, 162.2015.
10. Tasheva R. and Stefanov I,. 2020, proceedings of the Annual Scietific Conference of the National Millitary University “Vasil Levski”, 27 – 28 of May, 2021, Veliko Tarnovo, Bulgaria.
11. Stuchlik, Z., Kolos, M.: 2015, GReGr 47, 23.